\documentclass{article}
\usepackage{spconf,amsmath,graphicx,hyperref}
\usepackage{siunitx}

\title{\vspace{-2ex}Phase retrieval from 4-dimensional electron diffraction datasets}

\name{Thomas Friedrich\sthanks{\href{mailto:thomas.friedrich@uantwerpen.be}{\nolinkurl{thomas.friedrich@uantwerpen.be}}}, Chu-Ping Yu, Johan Verbeek, Timothy Pennycook, Sandra Van Aert}
\address{Electron Microscopy for Materials Science (EMAT) and NANOlab Center of Excellence \\ 
University of Antwerp, Groenenborgerlaan 171, 2020 Antwerp, Belgium}

\begin{document}
%
\begin{minipage}[t]{\textwidth}
\subsection*{Copyright notice}

\textcopyright 2021 IEEE. Personal use of this material is permitted. Permission from IEEE must be obtained for all other uses, in any current or future media, including reprinting/republishing this material for advertising or promotional purposes,creating new collective works, for resale or redistribution to servers or lists, or reuse of any copyrighted component of this work in other works
\end{minipage}

\maketitle

\begin{abstract}
We present a computational imaging mode for large scale electron microscopy data, which retrieves a complex wave from noisy/sparse intensity recordings using a deep learning approach and subsequently reconstructs an image of the specimen from the Convolutional Neural Network (CNN) predicted exit waves. We demonstrate that an appropriate forward model in combination with open data frameworks can be used to generate large synthetic datasets for training. In combination with augmenting the data with Poisson noise corresponding to varying dose-values, we effectively eliminate overfitting issues. The U-NET\cite{UNET} based architecture of the CNN is adapted to the task at hand and performs well while maintaining a relatively small size and fast performance. The validity of the approach is confirmed by comparing the reconstruction to well-established methods using simulated, as well as real electron microscopy data. The proposed method is shown to be effective particularly in the low dose range, evident by strong suppression of noise, good spatial resolution, and sensitivity to different atom types, enabling the simultaneous visualisation of light and heavy elements and making different atomic species distinguishable. Since the method acts on a very local scale and is comparatively fast it bears the potential to be used for near-real-time reconstruction during data acquisition. 

\end{abstract}
\begin{keywords}
phase retrieval, inverse problem, electron diffraction, 4D-STEM, CBED
\end{keywords}
\section{INTRODUCTION}
\label{sec:intro}
Transmission electron microscopy (TEM), and particularly scanning transmission electron microscopy (STEM) is one of the most powerful and versatile tools for material characterisation at the atomic scale. The increasing amount of characterisation techniques and the improving quality of collected data has been strongly linked to technological advancements, not only in the area of electron optics, but also in the field of detector systems \cite{Ophus2019}. 

Conventional electron detectors, such as annular bright field (ABF) or annular dark field (ADF), cover certain angular regions in reciprocal space and collect electrons scattered to corresponding scattering angles after their interaction with the specimen. By scanning the probe over the specimen, such a detector records a single integrated value for every position on the scan grid. 
This results in a 2-dimensional map of intensity values, from which we can deduce the positions and possibly also the types of the atoms under study \cite{bosch2015analysis}. 
The possibility of recording full, high-quality diffraction patterns in a reasonably short time is relatively recent and resulted in a whole new set of opportunities and challenges. Every scan point in a STEM experiment (2 spatial dimensions) contains a full diffraction pattern (2 dimensions in reciprocal space), thus providing a very large amount of information and potentially very large 4-dimensional (4D) datasets. One of the main challenges, dealing with this kind of data is the efficient and targeted extraction of information to translate a higher-dimensional dataset to a lower-dimensional one that is easy to interpret. Some existing methods include center-of-mass \cite{muller2014atomic} and integrated-center-of-mass \cite{lazic2016phase} imaging, as well as ptychographical methods, which can be computationally expensive \cite{OLeary2021,ePIE_MAIDEN20091256}, may discard some of the information present in the diffraction patterns \cite{SSB_RODENBURG1993304}, potentially amplify noise in low-dose experiments \cite{WDD_Rodenburg1992} or vary in their applicability due to constraining underlying model assumptions \cite{OLeary2021}.
In this paper, we present a new method of handling 4D datasets, which aims at constructing a phase image of the specimen. To have full flexibility in the number of scan points included in the dataset, we design the network to retrieve the phase and the amplitude of one convergent beam electron diffraction (CBED)-pattern at a time using only its neighbouring CBEDs, and then use the retrieved exit wave to reconstruct the local phase image of the material, employing the phase object approximation(POA). 
This process can be run already during the experiment, and the real-time reconstructed image therefore can help to validate appropriate settings and conditions of the microscope, which can reduce the chance of collecting unusable data, particularly at low-dose experiments. Furthermore, the method does not require to handle the entire dataset in memory or to perform repetitive read-write operations, making it computationally efficient and omitting common hardware limitations handling large datasets, while maintaining a high quality reconstruction, on par with other state of the art phase reconstruction algorithms, clearly outperforming conventional integrated-intensity-based imaging modes at delivering intuitively interpretable images. 

\section{MACHINE LEARNING}
\label{sec:ml}
Retrieving phase information from measured intensities is a classical inverse problem. It was shown that deep learning approaches can in principle be used to tackle these types of problems \cite{Ede,Laanait2019,Goy2018} in electron and light microscopy. The difficulty for 4D STEM arises from the fact that the data is typically rather noisy, implying that the recorded intensity does not directly correspond to the amplitude of the exit wave. Hence, to retrieve the exit wave means not only to solve the inverse problem to obtain the phase, but also to retrieve the wave amplitude from sparse intensity patterns. Since both amplitude and phase can generally be retrieved from the same set of adjacent diffraction patterns, we propose a single deep learning model to solve both problems simultaneously.  
\subsection{Training Data}
\label{sec:train_data}
\begin{figure}[hbt]
\begin{minipage}[b]{.48\linewidth}
  \centering
  \centerline{\includegraphics[ height=4.0cm]{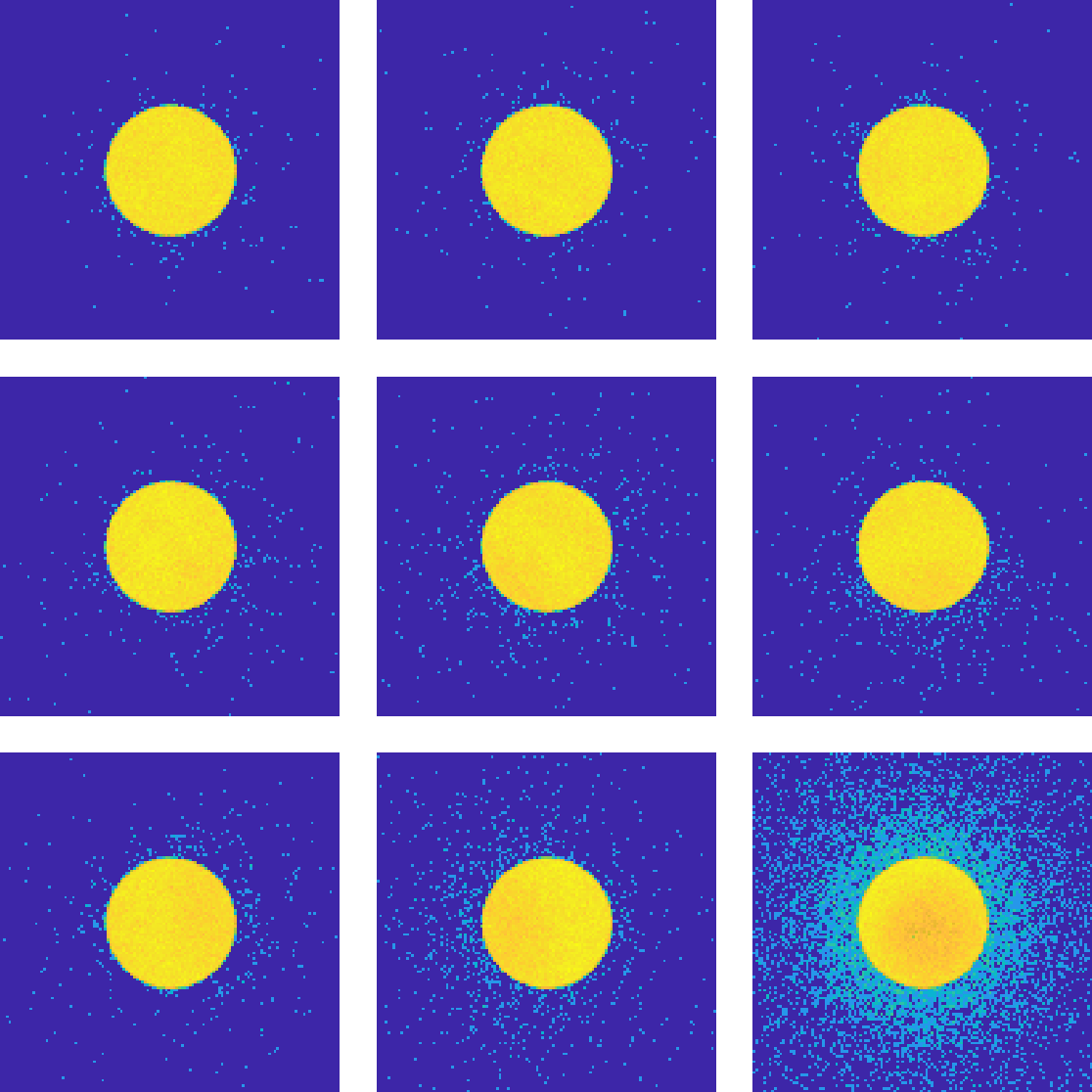}}
  \centerline{(a) Feature}\medskip
\end{minipage}
\hfill
\begin{minipage}[b]{0.48\linewidth}
  \centering
  \centerline{\includegraphics[height=4.0cm]{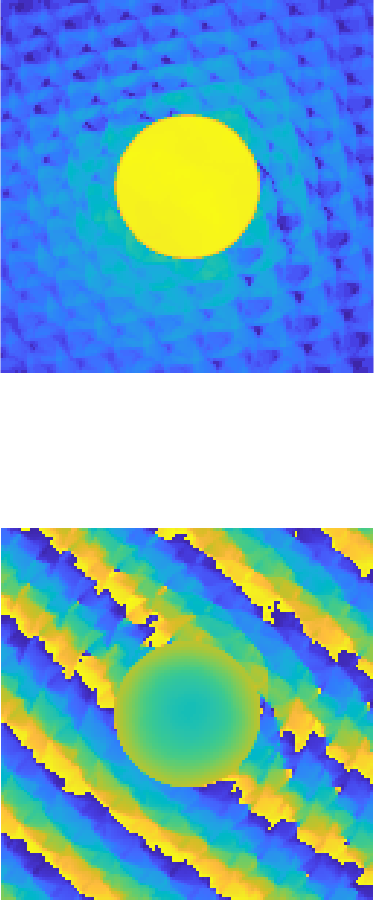}}
  \centerline{(b) Label}\medskip
\end{minipage}
\caption{Example of a feature-label set. (a) Feature: 3x3 kernel of neighbouring diffraction patterns.(log-scale) (b) Label: amplitude (top, log-scale) and phase (bottom) of the exit wave}
\label{fig:ex_feat_lab}
\end{figure}

We train our CNN with synthetic data, simulated using the multislice algorithm and microscope modelling as implemented in the MULTEM software \cite{Lobato2015}. This provides an appropriate forward model to compute electron probes for given microscope settings, its interaction with the electrostatic potential of atoms \cite{Lobato2014} and the resulting exit waves and diffraction pattern intensities. For computational efficiency we employed a relatively simplistic model, neglecting the effects of spatial and temporal incoherence and inelastic scattering. Each training sample consists of a 3x3 kernel of adjacent diffraction patterns as feature and an exit wave (amplitude-phase pair) as label in 128x128 pixels size as shown in figure \ref{fig:ex_feat_lab}. The simulation parameters and microscope settings are drawn at random from a uniform distribution within the limits of practically meaningful ranges. The atomic specimens for the simulations are generated from randomly drawn crystallographic data files ($\approx126000$) from the Materials project \cite{Jain2013}. The crystallographic orientation parallel to the electron beam propagation is drawn from the set of all low-index zone-axis orientations, while its rotation around the beam vector is random. The specimen thickness range is between \SI{2}{\angstrom} and \SI{10}{\angstrom}, strictly obeying the limits of the POA. The effect of a finite electron dose is modelled from the diffraction patterns assuming Poisson distributions of the electron counts on the detector pixels. The dose is applied as a factor scaling the simulated pattern and thus shifting the expected values of the Poisson distributions accordingly. This step is applied as a data augmentation step during the training, resulting in a different dose and dose realisation for each training sample in each epoch.
The combination of an appropriate forward model, a vast amount of structures, continuous microscope parameter ranges and an effective way of data augmentation enables the creation of very large datasets without redundancy and thus provides the means to train a neural network to solve the given problem in a very general manner at little risk of overfitting.  

\subsection{NEURAL NETWORK IMPLEMENTATION}
\label{sec:nn}
\begin{figure}[htb]
\centerline{\includegraphics[ width=1.0\linewidth]{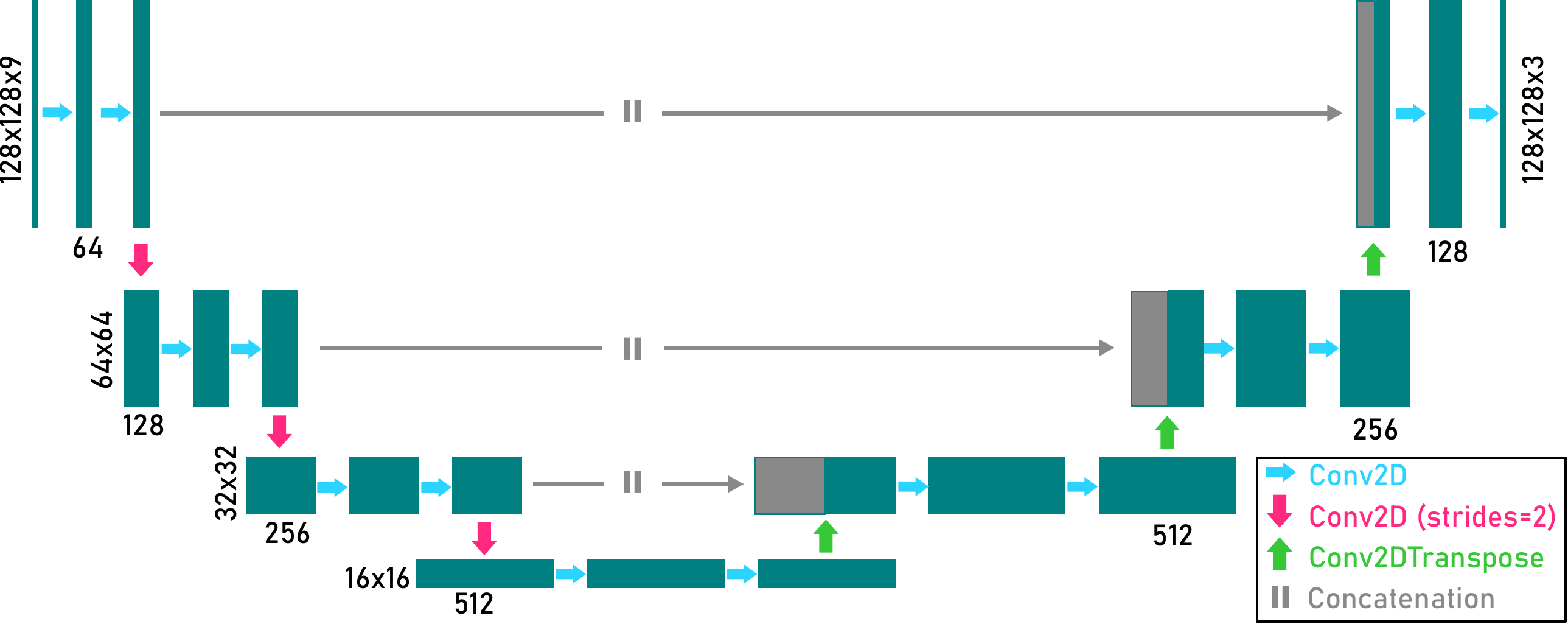}}
\caption{Network architecture}
\label{fig:unet}
\end{figure}

The network used is based on the U-NET architecture \cite{UNET} with some notable modifications as outlined in figure \ref{fig:unet}. The input is a stack of nine diffraction patterns of size 128x128 pixels. CBED patterns naturally exhibit a large difference between the bright-field and the dark-field regions. To exploit the valuable contribution of dark-field scattered electrons the input tensors intensity range is compressed by raising it to the power of 0.1, prior to standardising each input tensor by subtracting its mean and dividing by its standard deviation. Due to the small input size of the diffraction patterns in the spatial dimensions, this implementation of the U-Net goes only 3 level deep (instead of 4) to arrive at a minimum map size of 16x16 pixels. To avoid loss of information at the downsampling steps, convolutional layers with strides of two are substituted for the pooling layers, thus making the CNN fully convolutional. All convolutional layers use padding to maintain their map sizes.

\begin{figure}[!htb]
\begin{minipage}[b]{0.325\linewidth}
  \centering
  \centerline{\includegraphics{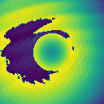}}
  \centerline{(a) wave phase $\phi$}\medskip
\end{minipage}
\begin{minipage}[b]{0.325\linewidth}
  \centering
  \centerline{\includegraphics{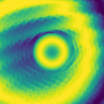}}
  \centerline{(b) $\sin{\left(\phi\right)}$}\medskip
\end{minipage}
\begin{minipage}[b]{0.325\linewidth}
  \centering
  \centerline{\includegraphics{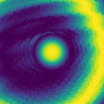}}
  \centerline{(c) $\cos{\left(\phi\right)}$}\medskip
\end{minipage}
\caption{Example of a phase decomposition.}
\label{fig:phase_step}
\end{figure}

Furthermore, leaky rectified linear units are used as activation functions, introducing a small slope on the negative part of the classical ReLU function. The output has three layers: the retrieved exit-wave amplitude and two layers representing the sine- and cosine components of the wave phase. The phase decomposition at the output avoids the necessity to estimate unnatural edges in the phase at the transitions between $\pi$ and $-\pi$ as shown in figure \ref{fig:phase_step}a, which have shown to hamper convergence and degrade reconstruction results. The outputs of these two layers are constrained by a scaled hyperbolic tangent function, forcing the output into a range between $\pm\pi$. Subsequently, their respective trigonometric functions are applied: $f_1(x) = \sin{\left(\tanh\left({x}\right)*\pi\right)}$ and $f_2(x) = \cos{\left(\tanh\left({x}\right)*\pi\right)}$. The output for the phase amplitude is simply linear. The Adam-optimizer is used to minimise the sum of the pixel-wise $\mathcal{L}_1$-losses and patch-wise structure dissimilarity-losses on the phase and amplitude images. We further penalise the euclidean norm of the decomposed phase tensor as it needs to be one for an accurate recombination.

\section{PHASE OBJECT RECONSTRUCTION}
In a microscope, electrons are emitted to interact with the atoms that compose the sample and the electric field they establish. By neglecting excitation events that reduce the coherency of the beam, the effect of the interaction can be seen as adding a phase shift to the coherent electron wave. Furthermore, if a thin sample is assumed so that the electron wave propagation can be omitted, then the interaction between the object and the electron can be described with the following equation:
\begin{equation}
O(\vec{r})\psi_{in}(\vec{r}) = \psi_{out}(\vec{r}).
\end{equation}
This is the phase object approximation, where \(O(\vec{r})\) describes the projected electrostatic potential or phase distribution of the object at each position \(\vec{r}\). \(\psi_{in}(\vec{r})\) and \(\psi_{out}(\vec{r})\) are the electron waves before and after the interaction with the object, respectively. From this equation we can see that if both \(\psi_{in}(\vec{r})\) and \(\psi_{out}(\vec{r})\) are perfectly known, the object is easily resolved. However, the fact is that neither of both can be directly measured in the microscope. One can make an assumption for the incident beam \(\psi_{in}(\vec{r})\) based on its projection to the reciprocal space \(\psi_{in}(\vec{k})\), since the latter is measurable and related to the former through Fourier transformation. Although the phase of the electron wave is lost when the intensity is recorded, we can assume a homogeneous phase distribution inside of the beam if the microscope is equipped with a probe corrector. The retrieval for the outgoing beam is much more complicated. After the scattering process both, phase and amplitude of its far field wave form \(\psi_{out}(\vec{k})\) do not have a regulated form anymore and thus a proper retrieval for both of them is needed.
There are multiple approaches to solve the phase, or even amplitude, retrieval problem. A list of methods that are categorised as ptychography \cite{rodenburg2004phase, jiang2018electron, song2019atomic} are among the most popular methods in electron microscopy. One of the most important criteria for ptychography reconstruction is the level of overlap between beam positions, connecting the partially retrieved object at one scan position with neighbouring CBEDs, and therefore the iterative process can eventually converge since the retrieved result from all scan positions agree with each other at their overlaps. In this sense, we design our neural network to predict the phase and the amplitude of the exit wave at each scan position based on the corresponding CBED and the CBEDs around it. Once the outgoing waveform for one CBED at the detector plane \(\psi_{out}(\vec{k})\) is retrieved, Fourier transformation is performed to acquire its counterpart at the object plane \(\psi_{out}(\vec{r})\). According to equation (1), the outgoing electron wave should contain a full description of the phase object. 
However, considering that detector regions delivering a weak signal bear very little information, the retrieved wave phase is weighted by the amplitude of the electron wave, which represents the confidence that the predicted phase values in reciprocal space are indeed reliable. This results in a weighted patch of the phase object at every scan point. By summing them together the phase image of the scan area can be reconstructed.
\label{sec:obj_rec}

\section{RESULTS}
\label{sec:results}
\begin{figure*}[htb]   %
\centering
    \begin{minipage}{0.735\textwidth}
      \includegraphics[width=\linewidth]{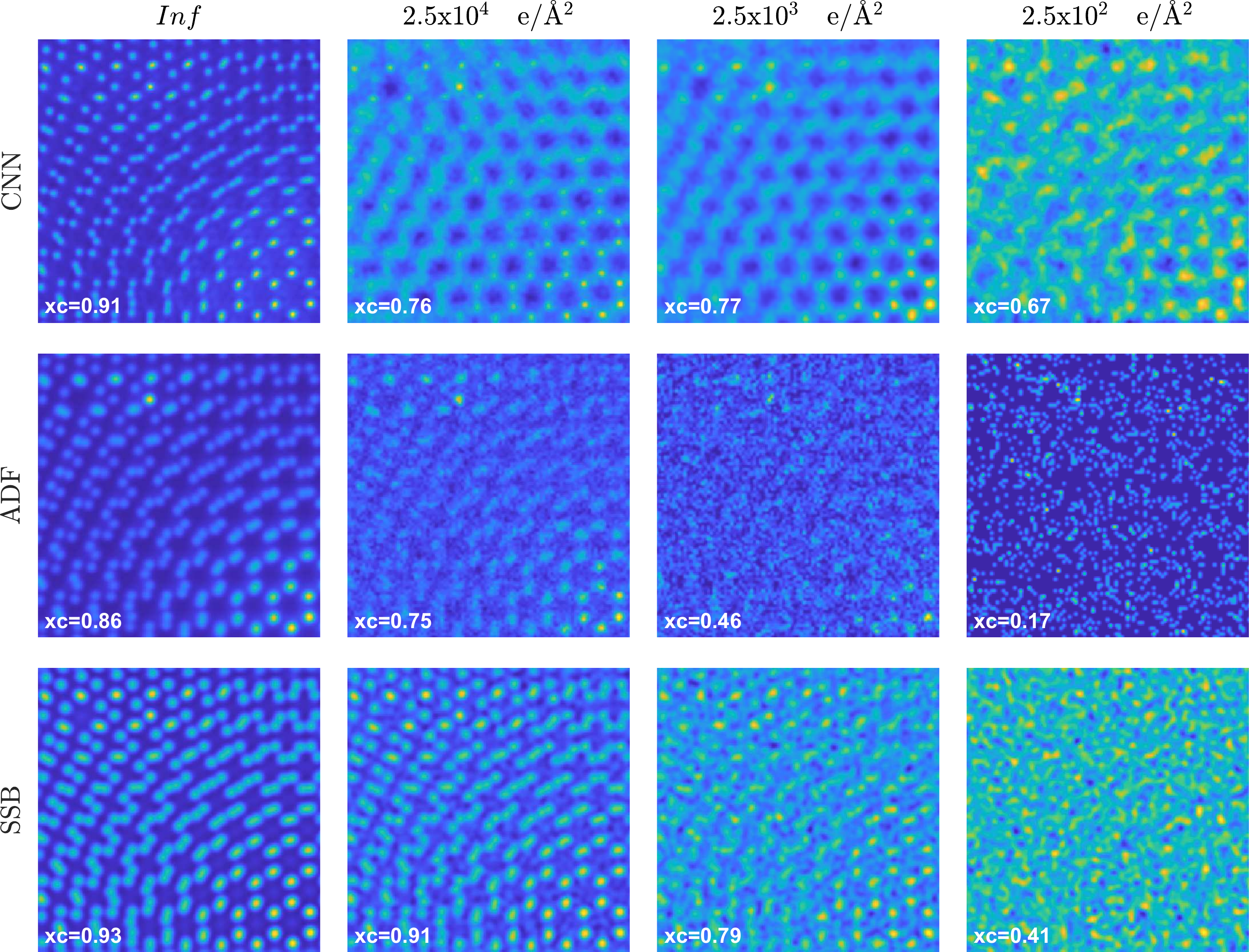}
    \end{minipage}
    \hfill
    \begin{minipage}{0.24\textwidth}
    \centering
        \includegraphics[width=0.8\linewidth]{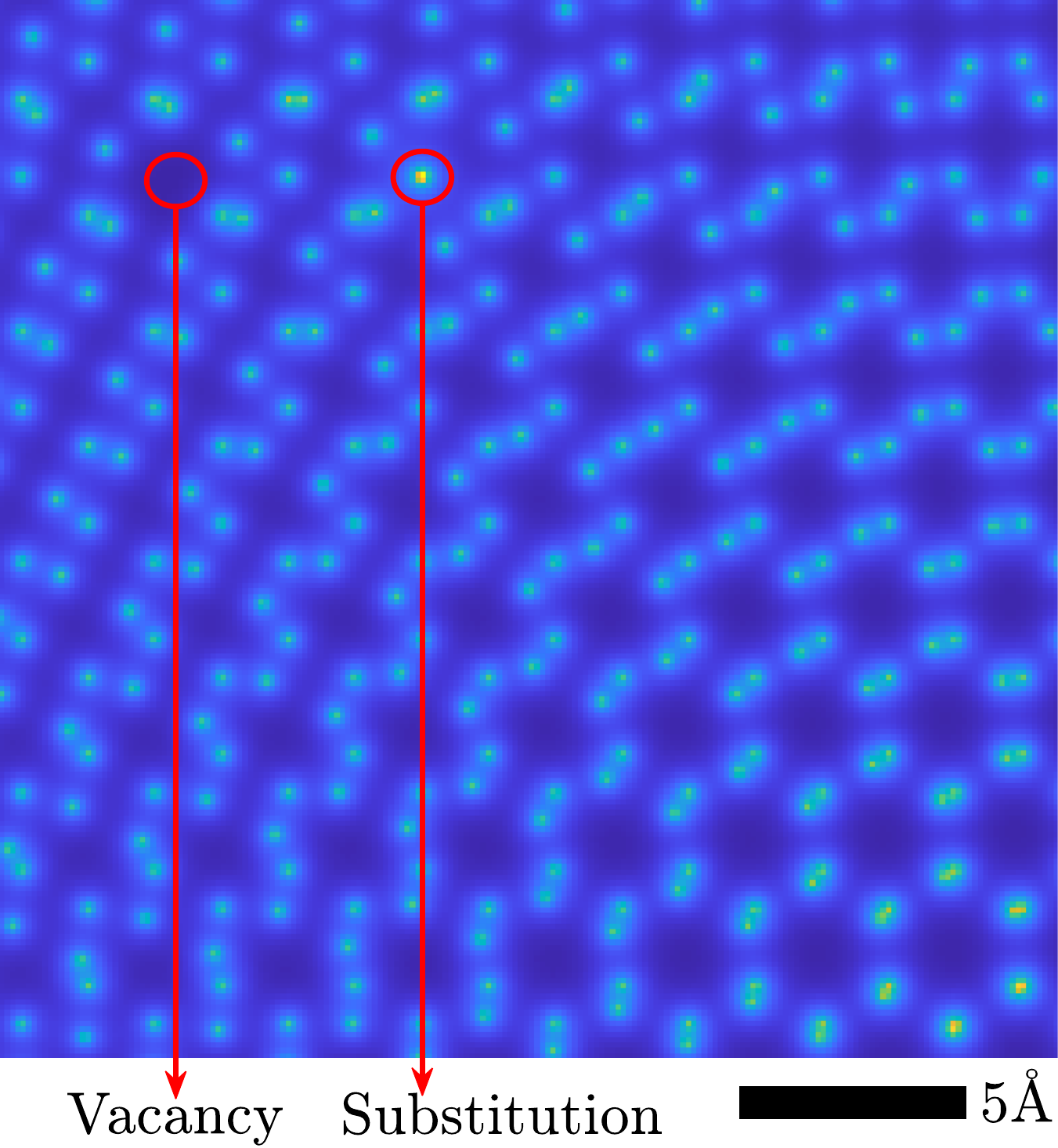} \\
        \caption{Left: Comparison of ADF imaging, SSB and the presented CNN reconstruction method for 4 dose settings. The images were retrieved from a simulated twisted bi-layer graphene dataset containing a vacancy and a Si substitution atom. Normalised cross-correlation values \textit{xc} are given w.r.t. the transmission function (top).}
    \end{minipage}
\label{fig:recon}
\end{figure*}

To test the performance of the neural network, we simulated 4D STEM datasets for reconstruction. The material of use is twisted bi-layer graphene, in which we deliberately created a vacancy, and also substituted one of the carbon atoms by a silicon atom. The simulation is performed under a convergent beam condition, with the convergence angle of 25~mrad, beam energy of 200~kV, and scanning step size of \SI{0.2}{\angstrom}. Poisson noise is added to the datasets to model the effect of a finite electron dose. The reconstruction is performed with the proposed method and the single side band ptychography reconstruction algorithm \cite{SSB_RODENBURG1993304} (SSB), and is compared with conventional ADF imaging. The results in figure \ref{fig:recon} show that the CNN reconstruction performs very well even in the presence of defects. Both the substitutional atom and vacancy can be clearly identified in the mid to high dose reconstructed image. The CNN reconstruction also shows a superior ability to retrieve a signal from low-dose data, as can be observed visually but is also indicated by the normalised cross-correlation values at \SI{2.5e2}{e/\angstrom^2} in figure \ref{fig:recon}.

\begin{figure}[hb]
\hfill%
\begin{minipage}[b]{0.41\linewidth}
  \centering
  \centerline{\includegraphics[width=0.9\linewidth]{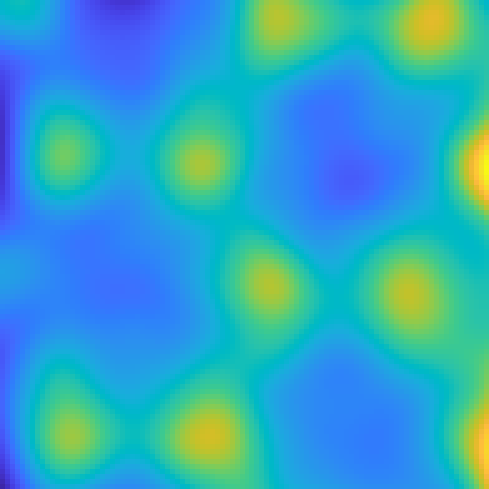}}
  \centerline{(a) Graphene}
\end{minipage}
\hfill%
\begin{minipage}[b]{0.41\linewidth}
  \centering
  \centerline{\includegraphics[width=0.9\linewidth]{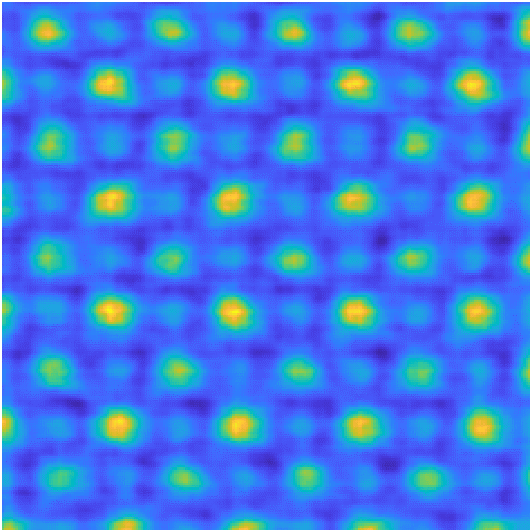}}
  \centerline{(b) $SrTiO_3$}
\end{minipage}
\hfill%
\hfill%
\caption{Reconstructions from experimental datasets.}
\label{fig:real_recon}
\end{figure}

We further tested the network with a real graphene dataset (Fig. \ref{fig:real_recon}a) using a convergence angle of 34 mrad, acceleration voltage of 60~keV, and scanning step size of \SI{0.04}{\angstrom}, and also a SrTiO$_3$ dataset (Fig. \ref{fig:real_recon}b) with 20~mrad convergence angle, 200~keV, and \SI{0.191}{\angstrom} step size as shown in figure \ref{fig:real_recon}. Strictly speaking, the latter is not a phase object due to its thickness, which leads to artefacts since the reconstruction should be done in a more complex manner. However, our result also shows that by rescaling the intensity of the dataset, it is still possible to generate a clear image that not only shows the positions of the atoms but also relative differences in the phase. However, the phase difference between the atoms and spacing does not match with the simulated transmission function of the specimen and therefore the reconstructed result is not yet suitable for further quantitative analysis.

\section{SUMMARY}
\label{sec:conclusion}
In this paper we demonstrate an imaging mode based on a two-step phase object reconstruction using deep learning to first address the phase- and amplitude retrieval problem for CBED patterns and secondly, reconstruct the phase object from the retrieved exit waves. The approach was validated on simulated, as well as real data, yielding comparable performance to state of the art conventional phase retrieval methods in terms of spatial resolution and sensitivity to different atomic species, particularly in the low-dose range for 2-dimensional and thin specimen.
A notable advantage of the method is that instead of going through iterative searching processes or post-processing sessions that can only be carried out after the whole dataset is acquired, our CNN can quickly predict phase and amplitude, and thus opens the possibility to perform reconstruction while collecting data nearly in real time at a quality 

\section{Acknowledgements}
We acknowledge funding from the European Research Council (ERC) under the European Union’s Horizon 2020 research and innovation programme (Grant Agreement No. 770887 PICOMETRICS and No. 802123 HDEM) and funding from the European Union’s Horizon 2020 research and innovation programme under grant agreement No. 823717 ESTEEM3. J.V. and S.V.A acknowledge funding from the University of Antwerp through a TOP BOF project. The direct electron detector (Merlin, Medipix3, Quantum Detectors) was funded by the Hercules fund from the Flemish Government.

\pagebreak

\bibliographystyle{IEEEbib}
\bibliography{refs}
\label{sec:ref}
\end{document}